\documentclass[amsmath,amssymb,showpacs,twocolumn]{revtex4}
\usepackage{graphicx}
\usepackage{bm}

\newcommand{\be}{\begin{eqnarray}}
\newcommand{\ee}{\end{eqnarray}}
\newcommand{\ve}{\varepsilon}

\newcommand{\ba}{\begin{array}}
\newcommand{\ea}{\end{array}}

\begin{document}
\title{
Teleportation of geometric structures in 3D
}

\author{Diederik Aerts $^1$, Marek Czachor $^{1,2}$, and {\L}ukasz Or{\l}owski $^{2}$}
\affiliation{
$^1$ Centrum Leo Apostel (CLEA) and Foundations of the Exact Sciences (FUND)\\
Vrije Universiteit Brussel, 1050 Brussels, Belgium\\
$^2$ Katedra Fizyki Teoretycznej i Informatyki Kwantowej\\
Politechnika Gda\'nska, 80-952 Gda\'nsk, Poland
}

\begin{abstract}
Simplest quantum teleportation algorithms can be represented in geometric terms in spaces of dimensions 3 (for real state-vectors) and 4 (for complex state-vectors). The geometric representation is based on geometric-algebra coding, a geometric alternative to the tensor-product coding typical of quantum mechanics. We discuss all the elementary ingredients of the geometric version of the algorithm: Geometric analogs of states and controlled Pauli gates. Fully geometric presentation is possible if one employs a nonstandard representation of directed magnitudes, formulated in terms of colors defined via stereographic projection of a color wheel, and not by means of directed volumes.
\end{abstract}

\maketitle

\section{Multivector geometry in 3D}

The fact that vector quantities can be interpreted geometrically in {\it at least\/} two different ways was clear already to H.~Grassmann \cite{G1844}, some 40 years before J. W. Gibbs \cite{Gibbs} and O. Heaviside \cite{Heaviside} invented vector calculus. One of the interpretations, close to what we are now accustomed to, treated vector $a$ as a directed line segment. Grassmann introduced the {\it outer product\/} $\wedge$ that allowed to extend two directed line segments into directed plane segments, or directed line and plane segments into directed volume segments (hence probably the name {\it linear extension theory\/} he gave to his formalism \cite{G1844}). The second Grassmann interpretation treated $a$ as a geometric point, $a\wedge b$ was a directed line segment determined by points $a$ and $b$, and $a\wedge b\wedge c$ was a directed plane segment determined by three points \cite{Hestenes}. In addition to the outer product $a\wedge b$ he introduced the inner product $a\cdot b$ acting, in a sense, in a way opposite to that of $a\wedge b$.

The two interpretations were not the only ones one could imagine. A variant of Grassmann's first interpretation (scalar and vector products) was used by Gibbs and Heaviside in their reformulation of Maxwell's electrodynamics. The two products are non-associative and define objects of different types (scalars $a\cdot b$ and pseudovectors $a\times b$, respectively), and any student knows one should not mix them with each other. It is interesting, however, that Grassmann himself {\it did\/} contemplate a combination $\lambda\, a\cdot b+\mu\, a\wedge b$, with arbitrary nonzero constants $\lambda$, $\mu$, and termed it the {\it central product\/}. It was W. K. Clifford who finally realized that the central product with $\lambda=\mu=1$ defines an operation which is indeed central to the algebra of vectors \cite{Clifford}. Clifford's {\it geometric product\/} $ab=a\cdot b+a\wedge b$ is associative and reconstructs the two products of Grassmann by $a\cdot b=\frac{1}{2}(ab+ba)$ and $a\wedge b=\frac{1}{2}(ab-ba)$.

The Grassmann-Clifford vector calculus is completely counterintuitive for all those who learned the Gibbs-Heaviside formalism at school, but there are reasons to believe that these were Gibbs and Heaviside who spoiled the work. Perhaps the most difficult conceptual element of the geometric product is that it mixes objects of apparently different species --- scalars and bivectors. But the problem is yet deeper since associativity allows to discuss products of arbitrary numbers of vectors, leading to combinations of all the four types of 3D objects --- scalars (directed points), vectors (directed line segments), bivectors (directed plane segments), and trivectors (directed volumes). Such general combinations are called polyvectors \cite{Pavsic} or multivectors \cite{Doran}.

Any directed line segment can be regarded as containing {\it two\/} types of directed objects of different dimensionality: The 1-dimensional interior and the 0-dimensional endpoints. The property is so obvious (``every stick has two ends") that does not, per se, deserve further comments. However, the subtlety we want to point out is that when it comes to the {\it directed magnitudes\/} themselves, it is by no means obvious that the interior should be equipped with the same directed value as the endpoints. The magnitude of the interior of a segment is typically identified with its {\it length\/}, and if we equip the segment with a kind of arrow we obtain an interpretation of its directed value. The procedure is no longer so natural if we turn to the endpoints, and thus in what follows we prefer to think of directed magnitudes in terms of {\it colors\/} (see below for a precise mathematical definition of what we mean by this statement).

The example of the 1D segment illustrates the first idea we will develop in this paper: Multivectors in 3D will be regarded as colored cubes of fixed (e.g. unit) size, whose interiors, walls, edges, and corners, have colors than can differ from one another. So the basic 3D shapes (cubic interiors, square walls, segments forming the edges, and the points where the edges meet) play the role of {\it blades\/} (Clifford geometric products of mutually orthonormal basis vectors) and the colors are the corresponding directed magnitudes. This type of geometric interpretation has an additional advantage of showing that a multivector is a single object whose different components are as inseparable from one another as the ends cannot be separated from the stick.

The second goal of this paper is to show that multivectors in 3D allow for geometric implementation of the quantum teleportation protocol \cite{Tel} entirely at the geometric level and without any reference to quantum mechanics. That formally it is possible is a trivial consequence of two facts. First, as shown recently in \cite{AC07,C07,AC08,MO,MP}, all quantum algorithms can be represented geometrically if one replaces $n$-bit entangled states from a $2^n$-dimensional complex Hilbert space by multivectors based on a Clifford algebra of some $n$-, $(n+1)$-, or $(n+2)$-dimensional (Euclidean or pseudo-Euclidean) space. Secondly, the simplest teleportation protocol is an example of a 3-bit quantum algorithm involving only real numbers. As such, it allows for a natural geometric representation in 3D, and thus is especially attractive from the point of view of geometric representations. Continuing in similar vein, one can extend the idea to a 3D lattice whose single cell is described by a single point, three edges, three walls, and one interior --- together $8=2^3$ basic elements typical of three dimensions --- but then one needs (at least) one more natural number to characterize the cell. The full algorithm involving complex amplitudes can be represented in geometric terms in 4D.

\section{Geometric-product coding}

Consider an $n$-dimensional real Euclidean space, and denote its orthonormal basis vectors by $b_k$, $1\leq k\leq n$. A normalized {\it blade\/} is defined by $b_{k_1\dots k_j}=b_{k_1}\dots b_{k_j}$, where $k_1<k_2<\dots <k_j$. The basis vectors (one-blades) satisfy Clifford's geometric algebra
\be
b_k\cdot b_l=\delta_{kl}=\frac{1}{2}(b_kb_l+b_lb_k).
\ee
The link between a binary number $A_1\dots A_n$ and blades (the $A$s are bits) is given by the formula
\be
c_{A_1\dots A_n}=b_1^{A_1}\dots b_n^{A_n}
\ee
where it is understood that $b_k^0=1$. The blades $c_{A_1\dots A_n}$ parametrized by binary sequences are occasionally referred to as combs.
Sometimes one needs complex numbers; their geometric-algebra analogs can be defined in several ways (cf. \cite{AC08}) but in the context of teleportation one deals with gates that are real, so for simplicity we skip this point.

Let $\psi$ be a general multivector in 3D,
\be
\psi
&=&
\sum_{A,B,C=0}^1\psi_{ABC}c_{ABC},
\ee
where $\psi_{ABC}$ are real numbers. Linking $\psi_{ABC}$ with colors by means of the stereographic projection of a color wheel \cite{wheel} shown in Fig.~1
\begin{figure}
\includegraphics[width=6 cm]{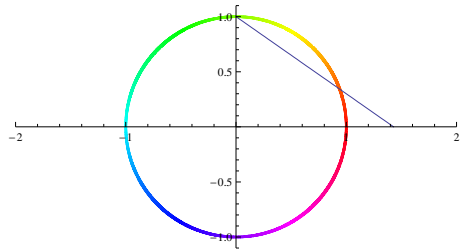}
\caption{Stereographic projection of a color wheel is a one-to-one map between real numbers and visible colors. If $x$ is a real number then the hue $h(x)$ of the color is computed in Mathematica according to $h(x)={\tt Hue}[\nu(x)]$, where $0\leq\nu(x)<1$ is defined implicitly by $x(1-\sin 2\pi\nu)=\cos 2\pi\nu$.}
\end{figure}
we obtain a geometric representation of $\psi$ whose special case is shown in Fig.~2.
\begin{figure}
\includegraphics[width=8 cm]{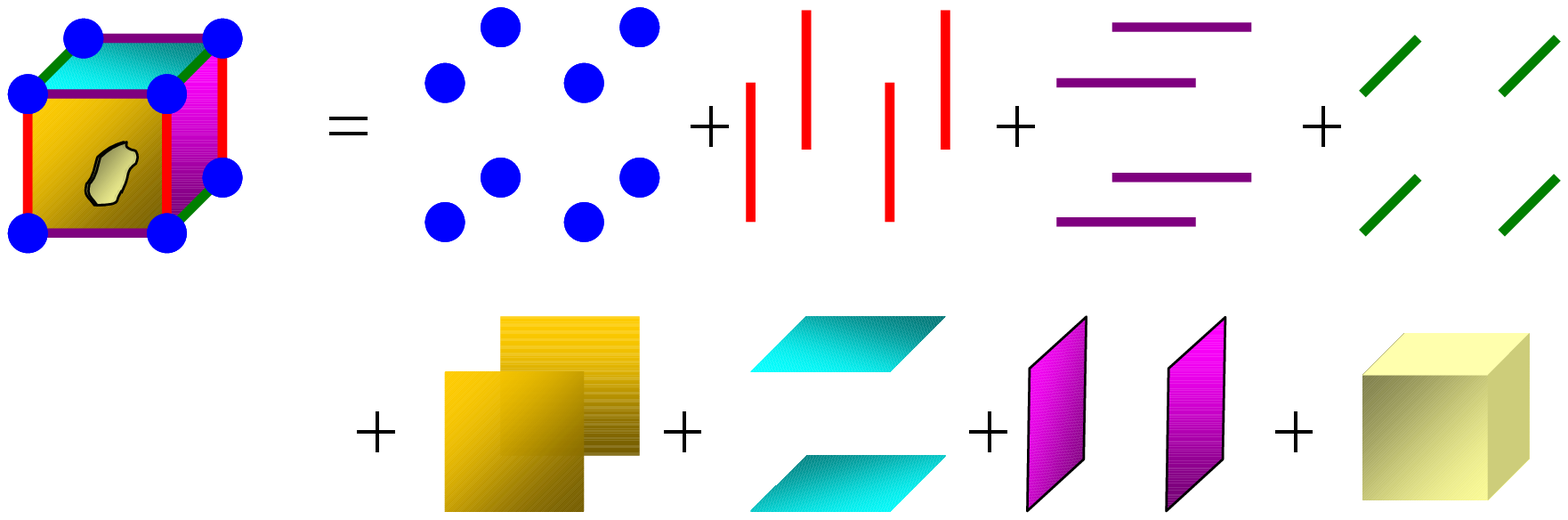}
\caption{Example of a general three-bit multivector $\psi=\sum_{ABC=0}^1 \psi_{ABC}c_{ABC}$. Values of the components $\psi_{ABC}$ can be deduced by means of the color wheel. Here we find approximately $\psi_{000}=-0.07$, $\psi_{100}=0.32$,  $\psi_{010}=-3.08$,  $\psi_{001}=1.06$, $\psi_{110}=-0.85$, $\psi_{101}=0.27$, $\psi_{011}=-0.86$,  $\psi_{111}=4.07$. Since $h(0)={\tt Hue}[3/4]$ the multivector should be, perhaps, shown on a dark-blue background corresponding to ${\tt Hue}[3/4]$, making invisible all elements with $x=0$ (but then the corners of the cube,  $\psi_{000}c_{000}$, would practically disappear from the figure). The representation is redundant in the sense that one could consider a single representative of each class: A blue point at only one corner (instead of 8), one red $z$-edge (instead of 4), one violet $z$--$y$-wall (instead of 2), and so on. Then the coloring method would be applicable to a cubic lattice, and not only to a single cube.}
\end{figure}

\section{Geometric gates and teleportation}

The teleportation protocol can be described in various ways, also in purely spacetime 2-spinor terms \cite{MCT}. The form which is especially useful here is the formulation in terms of a network of elementary gates acting on an initial state \cite{NC}.
In the standard quantum mechanical version one begins with the state
\be
|\psi_1\rangle
&=&
\alpha|0_1\rangle+\beta|1_1\rangle
\ee
which is to be teleported, and the entangled state
\be
|\Phi_{23}\rangle
&=&
\frac{1}{\sqrt{2}}
\Big(|0_20_3\rangle+|1_21_3\rangle\Big)
\ee
which plays a role of the carrier of quantum information, and is one of the four 2-bit entangled states forming the so-called Bell basis (Fig.~3 and 4). The Bell basis can be regarded as an analog of the Minkowski tetrad \cite{PR}, if one translates qubits into 2-spinors \cite{MCT} and $|\Phi_{23}\rangle$ is then an analog of the spacelike worldvector $x^a$ \cite{MCT}. The protocol does not need the concrete state $|\Phi_{23}\rangle$, but any non-factorizable two-bit state can be employed --- the 2-spinor protocols analyzed in \cite{MCT} employ analogs of $y^a$ and $\ve^{AB}$.

The goal is to implement the map
\be
|\psi_1\rangle
=
\alpha|0_1\rangle+\beta|1_1\rangle
\to
|\psi_3\rangle
=
\alpha|0_3\rangle+\beta|1_3\rangle
\ee
with unknown $\alpha$, $\beta$. The network of gates acts as follows
\be
H_1H_2Z{_3}{^1}X{_3}{^2}H_1X{_2}{^1}|\psi_1\rangle|\Phi_{23}\rangle
&=&
|0_10_2\rangle\big(\alpha|0_3\rangle+\beta|1_3\rangle\big),\nonumber\\
\ee
where $X_k$, $Z_k$, $H_k=(X_k+Z_k)/\sqrt{2}$ are the Pauli $X$ (the NOT gate) and $Z$, and Hadamard gates acting on $k$th bits; $X{_k}{^l}$, $Z{_k}{^l}$ are the Pauli gates acting on $k$th bits and controlled by $l$th bits. Below we shall give their explicit definition already in the geometric form, so let us first explain the geometric analog of teleportation. We begin with the multivectors
\be
\psi_1
&=&
\alpha c_{0_1}+\beta c_{1_1}=\alpha +\beta b_1,\\
\Phi_{23}
&=&
\frac{1}{\sqrt{2}}
\big(c_{0_20_3}+c_{1_21_3}\big)
=
\frac{1}{\sqrt{2}}
\big(1+b_{2}b_3\big).
\ee
\begin{figure}
\includegraphics[width=6 cm]{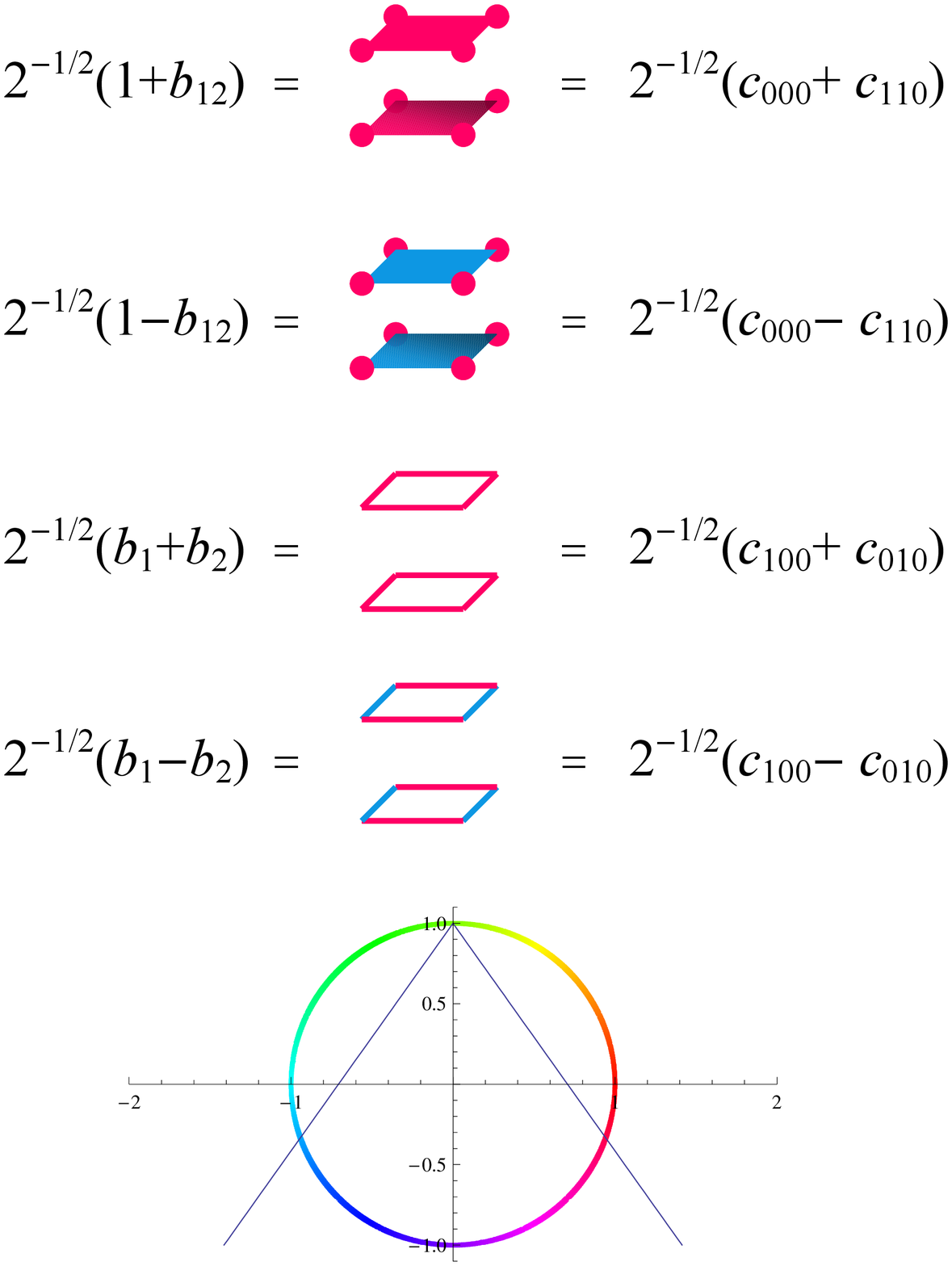}
\caption{Multivector analogues of the Bell basis of the first two bits. The color wheel is used to identify the colors corresponding to $\pm 1/\sqrt{2}\approx\pm 0.71$ (${\tt Hue}[0.554]$ and ${\tt Hue}[0.946]$).}
\end{figure}
The teleportation network must therefore act as follows
\be
H_1H_2Z{_3}{^1}X{_3}{^2}H_1X{_2}{^1}\psi_1\Phi_{23}
&=&
c_{0_10_2}\big(\alpha c_{0_3}+\beta c_{1_3}\big),\nonumber\\
&=&
\alpha +\beta b_3.
\ee
The elementary geometric gates act in direct analogy to their quantum counterparts. Below we list the nontrivial actions of the Pauli gates:
\be
X{_2}{^1}c_{100} &=& c_{110},\nonumber\\
X{_2}{^1}c_{101} &=& c_{111},\nonumber\\
X{_2}{^1}c_{110} &=& c_{100},\nonumber\\
X{_2}{^1}c_{111} &=& c_{101},\nonumber
\ee
\be
X{_3}{^2}c_{010} &=& c_{011},\nonumber\\
X{_3}{^2}c_{011} &=& c_{010},\nonumber\\
X{_3}{^2}c_{110} &=& c_{111},\nonumber\\
X{_3}{^2}c_{111} &=& c_{110},\nonumber
\ee
\begin{figure}
\includegraphics[width=5 cm]{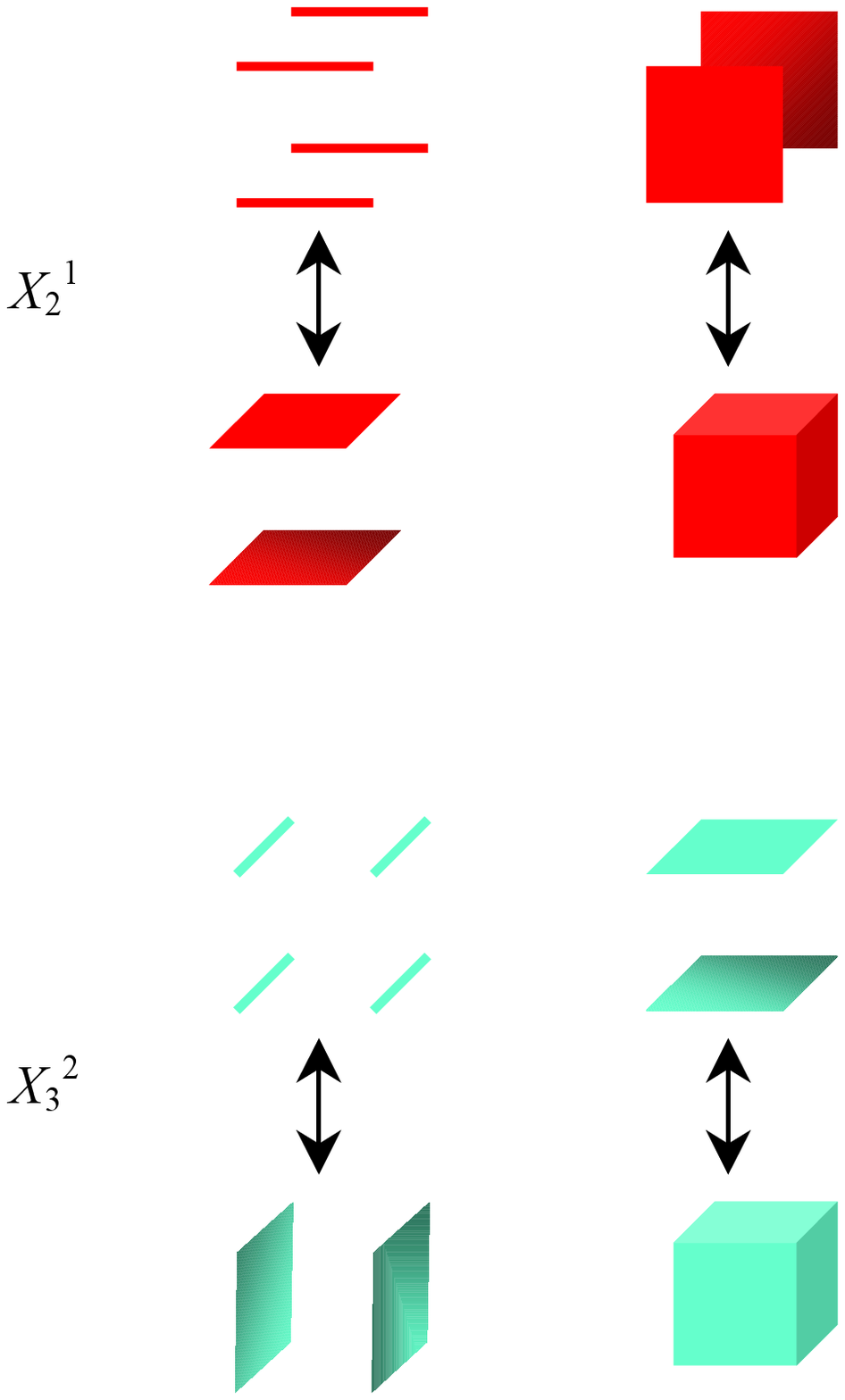}
\caption{Controlled $X$s. Only the blades containing $b_1$ are affected by $X{_2}{^1}$ ($b_1$ is the edge parallel to the $x$ axis, $b_{12}=b_1b_2$ is the unit square in the $x$-$y$ plane, and $b_{123}=b_1b_2b_3$ is the unit cube). Similarly, only the blades that contain $b_2$ are affected by $X{_3}{^2}$. The gates act trivially on the remaining blades.}
\end{figure}
\be
Z{_3}{^1}c_{100} &=& c_{100},\nonumber\\
Z{_3}{^1}c_{101} &=& -c_{101},\nonumber\\
Z{_3}{^1}c_{110} &=& c_{110},\nonumber\\
Z{_3}{^1}c_{111} &=& -c_{111},\nonumber
\ee
\begin{figure}
\includegraphics[width=8 cm]{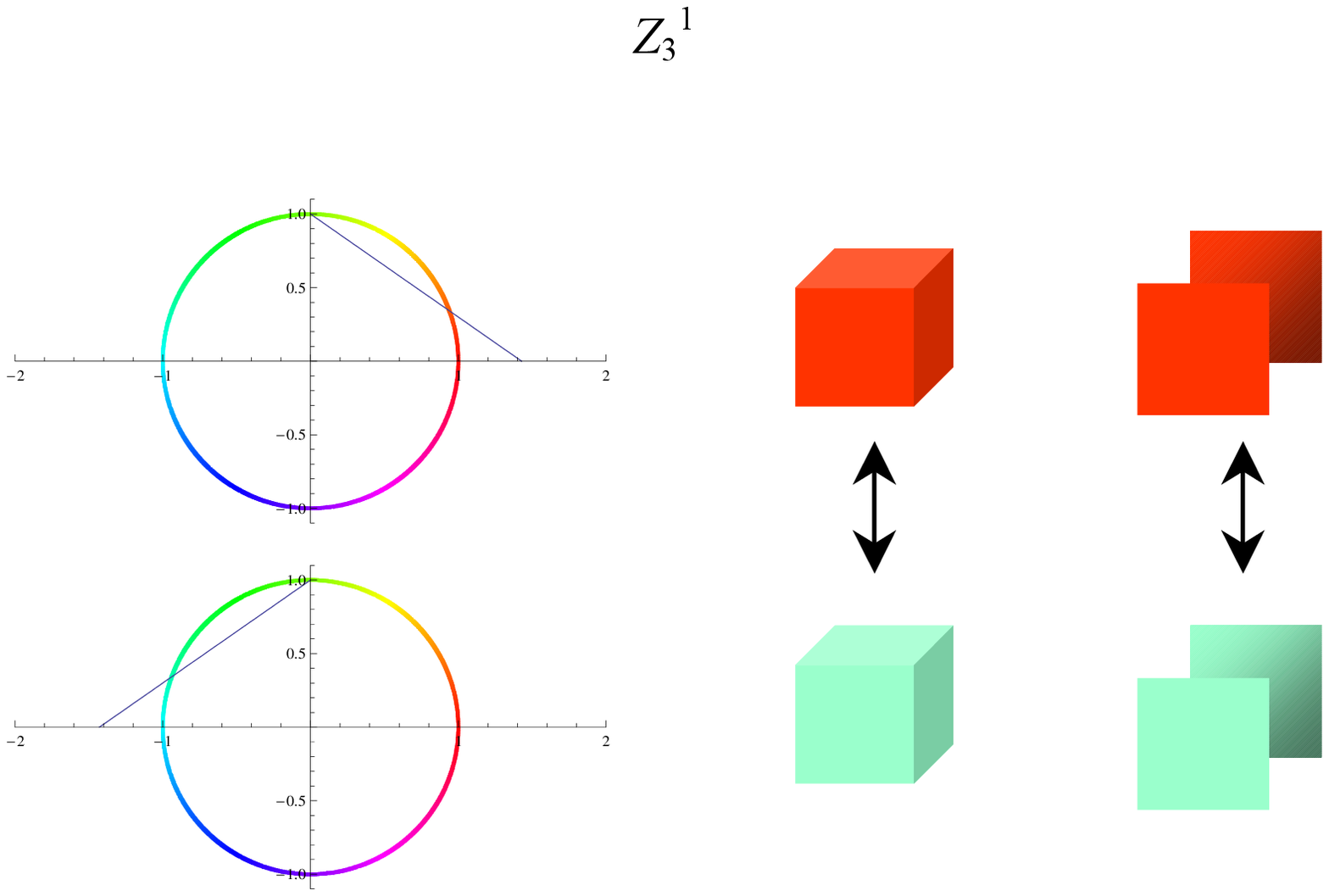}
\caption{$Z{_3}{^1}$ affects only those blades that contain $b_1$ (then the controlling first bit equals 1) and $b_3$. The gate changes color of the blade according to $h(x)\to h(-x)$. }
\end{figure}
\be
X{_1}c_{1BC} &=& c_{0BC},\nonumber\\
X{_1}c_{0BC} &=& c_{1BC},\nonumber\\
X{_2}c_{A1C} &=& c_{A0C},\nonumber\\
X{_2}c_{A0C} &=& c_{A1C},\nonumber
\ee
\be
Z{_1}c_{1BC} &=& -c_{1BC},\nonumber\\
Z{_2}c_{A1C} &=& -c_{A1C}.\nonumber
\ee
Translating these formulas into the language of blades we arrive at the following nontrivial actions of the controlled gates
\be
b_1 &\stackrel{X{_2}{^1}}{\leftrightarrow}& b_{12},\nonumber\\
b_{13} &\stackrel{X{_2}{^1}}{\leftrightarrow}& b_{123},\nonumber\\
b_2 &\stackrel{X{_3}{^2}}{\leftrightarrow}& b_{23},\nonumber\\
b_{12} &\stackrel{X{_3}{^2}}{\leftrightarrow}& b_{123},\nonumber\\
b_{13} &\stackrel{Z{_3}{^1}}{\leftrightarrow}& -b_{13},\nonumber\\
b_{123} &\stackrel{Z{_3}{^1}}{\leftrightarrow}& -b_{123}.\nonumber
\ee
The gates $X_k$ create or annihilate the basis vector $b_k$ in a blade (i.e. expand or squeeze the blade along the $k$th direction), and $Z_k$ change the sign of blade if $b_k$ is present (i.e. appropriately change the color of blades containing $b_k$).
Figures 5 and 6 show the geometry of the controlled Pauli gates. The Hadamard gates are a combination of the two actions. Fig.~7 shows the end result of the teleportation protocol.
\begin{figure}
\includegraphics[width=6 cm]{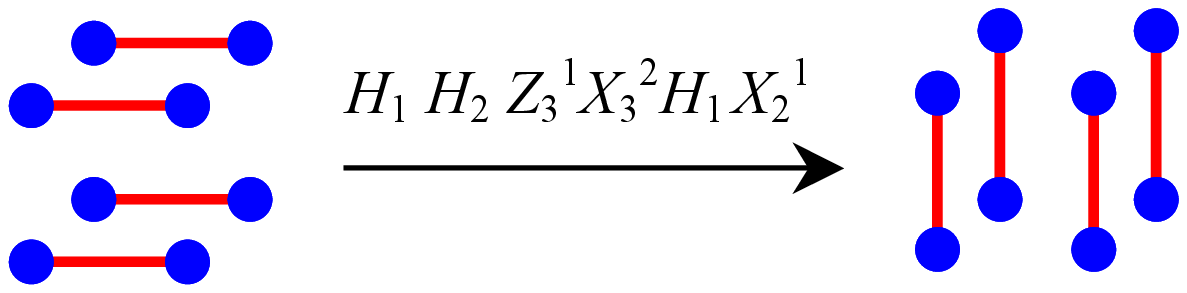}
\caption{The effect of the teleportation protocol on a multivector $\alpha +\beta b_1$.}
\end{figure}
\section{Representation on a deformed cubic lattice}

All the properties of our representation of multivectors are unchanged if one replaces the cubes by their color-preserving deformations. The lattice of such deformed cubes can be described by multivectors of the form
\be
\psi
&=&
\sum_{A,B,C=0}^1\sum_{N=0}^\infty\psi_{ABC,N}c_{ABC,N},
\ee
where $N$ labels different lattice cubes. 
If convenient, the natural number $N$ can be replaced by any $m$-tuple of natural numbers $(N_1,\dots,N_m)$, or a triple $(R_1,R_2,R_3)$ of real numbers indexing the ``center of mass" of the cell. An intuition behind this type of geometry is that the basis $b_1$,  $b_2$,  $b_3$, is associated with an internal degree of freedom analogous to the relative coordinate $\bm r$ occurring in 2-body problems, and $N$ is an analog of the center-of-mass coordinate $\bm R$.

Fig.~8 shows that the teleportation algorithm performs a discrete transformation between parts of the lattice, a kind of internal symmetry operation.

\begin{figure}
\includegraphics[width=9 cm]{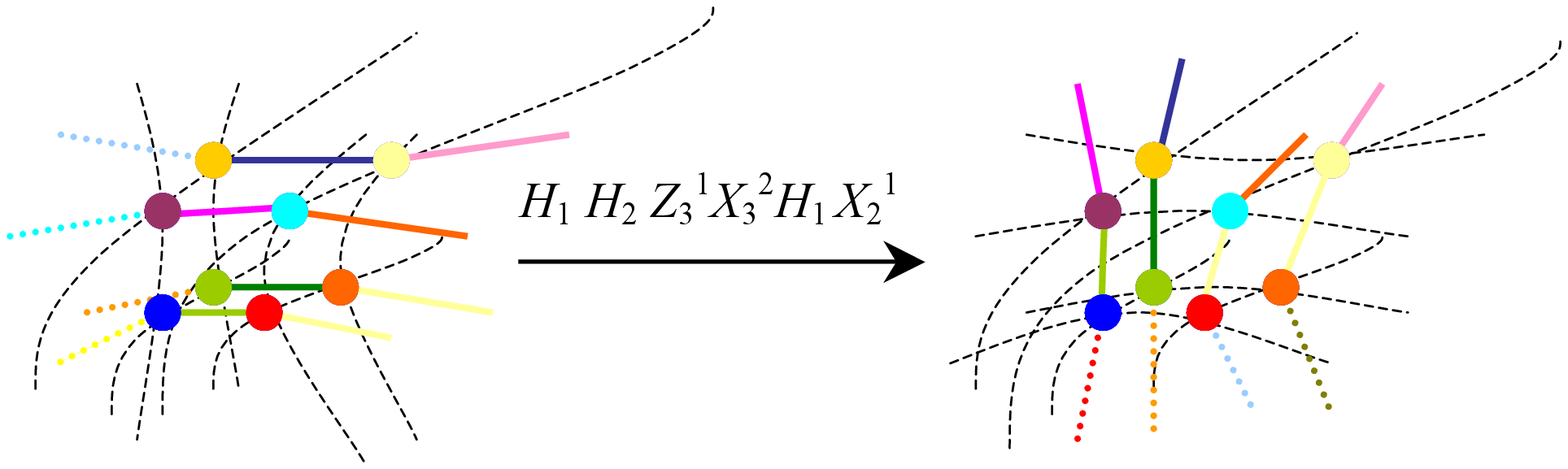}
\caption{The effect of the teleportation protocol on a general curved 3D lattice.}
\end{figure}

\section{Conclusions}

Algorithms that involve $k$-bit numbers must be based on geometric algebra of at least $k$-dimensional spaces ($k+1$ corresponds to problems that involve complex numbers) \cite{AC07,C07,AC08}. In effect, all 3- and 4-bit algorithms become  ``spacetime codes", to use the phrase of D. Finkelstein \cite{F1}. The teleportation protocol is a simple example from this class. Although in the present paper we work only with the geometric algebra of 3D Euclidian spaces, the transition to Minkowski space and more general Lorentzian manifolds is immediate \cite{H1}.

A natural geometric arena for geometric analogs of quantum teleportation is provided by 3D or 4D lattices, whose basic cells can be regarded as multivectors of dimension $2^3$ or $2^4$, respectively. In this context one should mention two immediate associations with earlier works. First of all, the lattice structure might be inherited from the lattices one finds in spin foam models \cite{Perez}. The second straightforward link is the idea of field theory defined on the Clifford space of points, areas and volumes \cite{Pavsic,Pavsic1}. In all these approaches the basic geometric intuitions are similar to what we have described above. What is novel in our approach is the possibility of quantum-like coding directly at the geometric level, with no need of quantization of any sort. It is quite remarkable that the celebrated teleportation algorithm fits into various spacetime structures in so natural way.


\begin{references}
\bibitem{G1844}H. Grassmann, {\it A New Branch of Mathematics: The Ausdehnungslehre of 1844 and other works\/}, translated by L. C. Kannenberg
(Open Court, Illinois, 1995)
\bibitem{Gibbs}J. W. Gibbs, {\it The  Scientific Papers of J. Willard Gibbs\/} (Longmas, Green and Company, London, 1906).
\bibitem{Heaviside}O. Heaviside, {\it Electromagnetic Theory\/} (Dover Publications, New York, 1950).
\bibitem{Hestenes}D. Hestenes, Grassmann's vision, in {\it Hermann Gunther Grassmann (1809-1877): Visionary Mathematician, Scientist and Neohumanist Scholar\/}, Gert Schubring, Ed. (Kluwer Academic Publishers, Dordrecht, 1996).
\bibitem{Clifford}W. K. Clifford, Applications of Grassmann's extensive algebra, American Journal of Mathematics Pure and Applied {\bf 1}, 350--358 (1878).
\bibitem{Pavsic}
M. Pav\v{s}i\v{c}, {\it The Landscape of Theoretical Physics: A
Global View. From Point Particles to the Brane World and Beyond, in Search of a Unifying Principle\/} (Kluwer, Boston, 2001).
\bibitem{Doran}C. Doran and A. Lasenby, {\it Geometric Algebra for Physicists\/} (Cambridge University Press, Cambridge, 2003).
\bibitem{Tel}C. H. Bennett, G. Brassard, C. Cr\'epau, R. Jozsa, A. Peres, and W. K. Wooters, 	
Teleporting an unknown quantum state via dual classical and Einstein-Podolsky-Rosen channels, Phys. Rev. Lett. {\bf 70}, 1895 (1993).
\bibitem{AC07}D. Aerts and M. Czachor, Cartoon computation: Quantum-like algorithms without quantum mechanics, J. Phys. A {\bf 40}, F259 (2007).
\bibitem{C07}M. Czachor, Elementary gates for cartoon computation, J. Phys. A {\bf 40}, F753 (2007).
\bibitem{AC08}D. Aerts and M. Czachor, Tensor-product versus geometric-product coding, Phys. Rev. A {\bf 77}, 012316 (2008).
\bibitem{MO}T. Magulski and {\L}. Or{\l}owski, Geometric-algebra quantum-like algorithms: Simon's algorithm, preprint arXiv:0705.4289 [quant-ph].
\bibitem{MP}M. Paw{\l}owski, Superfast algorithms and the halting problem in geometric algebra, preprint quant-ph/0611051.
\bibitem{wheel}A circle representation of colors was introduced already by Isaac Newton (Newton's color wheel) in his {\it Optics\/} (1706). Other color wheels are associated with the names of Hoener, Munsell, and Ostwald, cf. P. Zelansky and M. P. Fisher, {\it Design Principles and Problems\/} (Thomson Learning, 1996). We employ the {\it hue color wheel\/}, discussed in detail in D. Briggs, {\it The Dimensions of Colour\/}, http://www.huevaluechroma.com.
\bibitem{MCT}M. Czachor, Teleportation seen from spacetime: on 2-spinor aspects of quantum information processing, Class. Quantum Grav. {\bf 25}, 205003 (2008), preprint arXiv:0803.3289 [quant-ph].
\bibitem{NC}M. A. Nielsen, I. L. Chuang, {\it Quantum Computation and Quantum Information\/} (Cambridge University Press, Cambridge, 2000).
\bibitem{PR}R. Penrose and W. Rindler, {\it Spinors and Space-Time, vol.
1: Two-Spinor Calculus and Relativistic Fields\/} (Cambridge University Press, Cambridge, 1984).
\bibitem{F1}D. Finkelstein,	Space-time code, Phys. Rev. {\bf 184}, 1261 (1969).
\bibitem{H1}D. Hestenes, {\it Space-Time Algebra\/} (Gordon and Breach, New York, 1966).
\bibitem{Perez}A. Perez, The spin foam representation of loop quantum gravity, in {\it Approaches to Quantum Gravity. Toward a New Understanding of Space, Time and Matter\/}, ed. by  D.~Oriti  (Cambridge University Press, Cambridge, 2008), gr-qc/0601095.
\bibitem{Pavsic1}M.~Pav\v{s}i\v{c}, An extra structure of spacetime: A space of points, areas and volumes, Talk presented at the XXIX Spanish Relativity Meeting ERE 2006, 4th-8th September 2006, Palma de Mallorca, Spain, preprint gr-qc/0611050.
\end{references}
\end{document}